# Towards Transportation Digital Twin Systems for Traffic Safety and Mobility Applications: A Review

Muhammad Sami Irfan, Sagar Dasgupta, *Student Member, IEEE*, and Mizanur Rahman, *Member*

*Abstract*— Digital twin (DT) systems aim to create virtual replicas of physical objects that are updated in real time with their physical counterparts and evolve alongside the physical assets throughout its lifecycle. Transportation systems are poised to significantly benefit from this new paradigm. In particular, DT technology can augment the capabilities of intelligent transportation systems. However, the development and deployment of networkwide transportation DT systems need to take into consideration the scale and dynamic nature of future connected and automated transportation systems. Motivated by the need of understanding the requirements and challenges involved in developing and implementing such systems, this paper proposes a hierarchical concept for a Transportation DT (TDT) system starting from individual transportation assets and building up to the entire networkwide TDT. A reference architecture is proposed for TDT systems that could be used as a guide in developing TDT systems at any scale within the presented hierarchical concept. In addition, several use cases are presented based upon the reference architecture which illustrate the utility of a TDT system from transportation safety, mobility and environmental applications perspective. This is followed by a review of current studies in the domain of TDT systems. Finally, the critical challenges and promising future research directions in TDT are discussed to overcome existing barriers to realize a safe and operationally efficient connected and automated transportation systems.

## I. INTRODUCTION

The increasing trend towards connectivity and automation in the realm of transportation has spurred interest in the adoption of technologies for better management of traffic. The advent of connectivity has enabled the collection and aggregation of data from transportation systems at all levels, including mobile and stationary elements. Connected and Automated Vehicles (CAVs) can transmit and share vehicle information with roadside infrastructure and other vehicles. Similarly, connected roadside infrastructure can transmit information about the current road conditions and traffic states. Based on these connected systems, a number of service packages have been developed that provide specific mobility related services and benefits [1].

Despite the existence of these connected systems, the full potential of the massive data collection remains untapped. In this regard, Digital Twin (DT) technology promises to take advantage of this connected ecosystem and provide tangible benefits to the transportation system. DT has been defined as "a virtual representation of what has been produced"[2]. Essentially, a DT creates a digital replica of a physical asset by utilizing the real-time states of that object. It is important to note that a requirement of DT representation of an object is that there must be real time synchronization of the virtual and physical object through a frequent update of the state information. DT systems are often considered analogous to the Internet of Things (IoT) or Cyber-Physical Systems (CPS). While DT systems share components with IoT and CPS, there is a fundamental difference between these systems. DT systems are broader in scope and consideration and encompass IoT and CPS, but the same is not true the other way around. IoT systems comprise of interconnected physical devices with embedded sensors and processing capabilities [3]. On the other hand, CPS systems connect physical objects with the cyber world and integrate feedback action based on the input from the physical objects so that the system can provide some output in the form of decision or actuation [4]. However, both IoT and CPS systems lack the concept of maintaining a virtual twin of the physical asset. As the virtual twin is intended to be a direct representation of the physical object, it can be used to perform predictive analytics that would otherwise be impossible to perform on the physical asset itself.

DT systems are finding applications across a wide range of industries [5]. In manufacturing, DT systems have been used for applications such as maintenance, product lifecycle management and production planning control [6]. The aviation industry has found use of digital twins for design, assembly, manufacturing, operation and maintenance of aircrafts [7]. Power systems also benefit from the adoption of DT technologies. Power system digital twins offer power equipment malfunction detection, load prediction, intelligent image inspection, power system analysis, and health state evaluation [8]. In transportation, the proliferation of Intelligent Transportation System (ITS) technology has seen the widespread deployment of sensors and connectivity across the entire transportation ecosystem. This has made it possible to adopt DT technology in the management of traffic. A DT system in the context of transportation is a real-time digital representation of a physical transportation asset or process. Indeed, prototype systems targeting specific use cases are being developed to realize the benefits of a Transportation Digital Twin (TDT) system. Applications such as cooperative ramp merging [9], cooperative driving [10], and driver behavior modeling [11], [12] have been developed to illustrate the utility of TDT systems in improving the safety and mobility of existing transportation systems.

Even in the backdrop of such interest in TDT systems, there has been no consensus in terms of developing a unified framework and architecture for the realization of TDT systems. The majority of the implementations have developed architectures that are specific to their use cases. However, in order to ensure cross-compatibility across different TDT systems, it is essential to have an overarching reference architecture that can guide practitioners in developing TDT systems for any particular use case and at any particular level of transportation systems ranging from individual vehicles to city wide DT systems of transportation networks.

In developing a TDT that is focused on safety, mobility and environmental applications, it is required to have a system that has predictive capabilities such that actions can be taken proactively in order to avoid safety, mobility and environmental issues from occurring in the first place. Transportation systems, in particular, are a challenging field to implement DT systems on. A myriad of challenges, such as enormous scale, heterogenous data types, cost of implementation, etc., present significant roadblocks to their design and deployment. Thus, the focus of this paper is to address the issue of implementing TDT systems by reviewing the existing literature on transportation-related DT systems, presenting a reference architecture and framework for TDT systems focused upon safety, mobility and environmental aspects, and identifying the challenges that arise from the requirements of such TDT systems.

## II. TRANSPORTATION DIGITAL TWIN

The concept of DT in transportation systems is still in the nascent stage and consequently a formal definition of a TDT system has yet to be formulated. Based upon the general definitions of DT systems, a TDT system may be defined as a virtual representation of transportation systems that maintain a digital replica of physical transportation system elements such as vehicles, roadways, pedestrians etc. A TDT system must be a real time representation of the corresponding physical transportation elements to mimic the properties, behavior, and interactions between various transportation elements.

The key motivations behind implementing TDT are to improve safety and mobility and reduce greenhouse gas emissions of transportation systems. Therefore, implementing a TDT will be justified if it can provide benefits specific to these critical areas. Although implementing a DT of any system involves a series of stages, it is possible to classify them into two distinct phases: pre-digital twin and post-digital twin phases. These two phases are classified according to the stage of development of the DT system, whereby pre-digital twin refers to the system prior to actual implementation and the post-digital twin refers to the system after it has been implemented and deployed. Pre-digital twin phase consists of the following associated stages [13]:

1. *Pre-digital twin:* This is the initial digital prototype that is built to model a physical object. At this stage, the physical object does not exist; hence, no data coupling can be found. Pre-digital twins can be used to diagnose any technical issues in the digital vesion of an asset before its corresponding physical unit is built. An example of a pre-digital twin would be a digital prototype of a bridge section before it is actually constructed.

2. *Digital model:* A digital model extends the pre-digital twin by including the physical object as well. However, in this model there is no automatic data transfer between the digital and the physical objects. Once the bridge is built, it can be linked with its digital prototype to obtain a digital model of the bridge. However, within this model there is no automatic data transfer. Data collected by the sensors on the physical bridge needs to be manually fed into the digital prototype to realize the digital model of the bridge.

3. *Digital shadow:* In this stage, there is automated unidirectional data flow from the physical object to its digital counterpart. It also keeps data records of the physical object over time. This enables and facilitates future analysis. By automating the process of data transfer from the physical bridge asset to its digital version, a digital shadow of the bridge can be created. The digital shadow of the bridge will retain the historical states of the bridge data and can be used to assess the historic state of the bridge. However, no data flow occurs from the digital version of the bridge to the physical bridge asset.

In the post-digital twin phase, the following stages can be identified:

1. *Digital twin:* Here there is automated bidirectional data transfer between the digital and physical objects in real time. The DT is a real-time representation of its associated physical object. Data from the physical object is used to update the DT and control command from the DT that can be transmitted to the physical object to change its state. Continuing the example of the bridge, once bidirectional data flow exits between the physical and digital versions, the system can be deemed a DT of the bridge. The data flow from the digital version to the physical bridge can be advisory messages for the driver in the form of Variable Message Sign (VMS) updates.

2. *Digital sibling:* This is the most advanced stage of a DT whereby in the digital world, additional versions of the DT are created. The DT representation maintains an exact replica of a physical object and changes state only if the state of the actual object changes. On the other hand, the digital siblings aim to perform predictive analytics by constructing parallel simulated versions of the DT based on different cases and criteria. In this way, if one parameter in the physical world is altered, the effect on the state of the physical world and associated objects can be predicted. Hence when data analytics and simulations are used to predict the future structural health of the bridge, it can be considered a digital sibling of the physical bridge.

In order to facilitate the discussion of TDT during development, it useful to have a scheme of categorization that can help to describe a TDT system. For this purpose, the following classification is presented here [13], [14]:

1. *Lifecycle-based classification:*
   a. Digital Twin Prototype (DTP): The DT prototype is a digital version of a physical prototype that can be used to conduct tests that are not otherwise possible with physical prototyping. For example, a vehicle DTP may be used to model the dynamics of a specific type of vehicle. The vehicle DTP can then be used to run a large number of simulations to assess the crash safety of the vehicle. However, doing the same in the real world using a physical vehicle would not be feasible due to constraints of resources such as time and funding.
   b. Digital Twin Instance (DTI): Once the physical component of the DT is produced, it becomes a DT instance. DTI results when the DTP is interfaced with its physical twin version after the physical asset is manufactured. Each additional unit of the physical twin will result in an additional DTI.

2. Value proposition-based classification:
   a. *Supervisory Digital Twin (SDT):* SDT constitute the most basic versions of digital twins and are limited to representing the state of the physical twin or process that is being replicated. These systems simply present information that the related personnel can use in making decisions and taking action. In a transportation context, an example of an SDT system is a real-time traffic visualization system for a roadway section. Such a system provides a real-time visualization of the traffic situation through real-time video feeds and does not provide any interaction capability or predictive features.
   b. *Interactive Digital Twin (IDT):* An interactive digital twin, in addition to representing real time state of the physical object, is also able to control some aspects of those objects. By making control decision based on real time state information, an IDT is able to achieve better control than a human operator taking decisions based on information from an SDT. The real time traffic visualization system described above can be considered an IDT if the system integrates the ability to interact with the road vehicles. Such interactions can be in the form of advisory messages sent to VMS signs along the road.
   c. *Predictive Digital Twin (PDT):* This type is the most complex implementation of a digital twin. PDT systems are able to monitor the state of the physical object and use sophisticated algorithms and modelling to achieve proactive capabilities. An example of a PDT system would be a traffic rerouting application that optimizes the travel times for all vehicles. Such a system not only monitors the traffic, but also performs extensive computations to predict the future state of the traffic and optimize accordingly. Furthermore, it is able to send out rerouting advisory information to the vehicles.

## III. REFERENCE ARCHITECTURE FOR TRANSPORTATION DIGITAL TWIN

The decision to develop and implement a TDT system will first need to be motivated by the right justifications. In this regard, the following three considerations have been identified as the most decisive factors in building such systems: Complexity, Breadth and Depth [15]. Complexity refers to the cost of producing the DT system, in terms of time as well as money. Breadth of the DT will be determined by the requirement related to the generality or specificity of the DT system. Finally, the depth of the DT system will be determined by the precision and accuracy requirement of the system to be designed and implemented.

In the context of TDT systems as well, these considerations will manifest into the design and implementation process. Thus, a hierarchical concept of the TDT is presented in this section to address these considerations. Figure 1 illustrates the proposed hierarchical concept. The concept builds from the least level of complexity and breadth up to the most. The lowest level within the hierarchy is the individual asset twins whereby the DT representation is only built for individual discrete units such as a single vehicle, a ORU or a traffic signal controller. At this level, the DT systems will have low complexity due to limited interaction with other elements of the transportation systems. The breadth of the individual asset twin will be very narrow as they may be very specific to a vehicle model or individual ORU. However, in contrary to the other considerations, the depth requirement at this level may actually be higher. Indeed, for building any applications using the DT of the individual assets, it would be advantageous to have highly accurate and precise data.

Immediately above the individual asset twin level is the intersection level twin, which incorporates the individual assets as well the DT representation of an intersection. Thus, on top of the elements of the individual asset twins, this level will contain the intersection geometry information, the state of the traffic at the intersection at any instant as well information of any incidents at that intersection. At this level, the interaction between the individual elements starts to become relevant and it is important to model this interaction in the digital space of the TDT system.

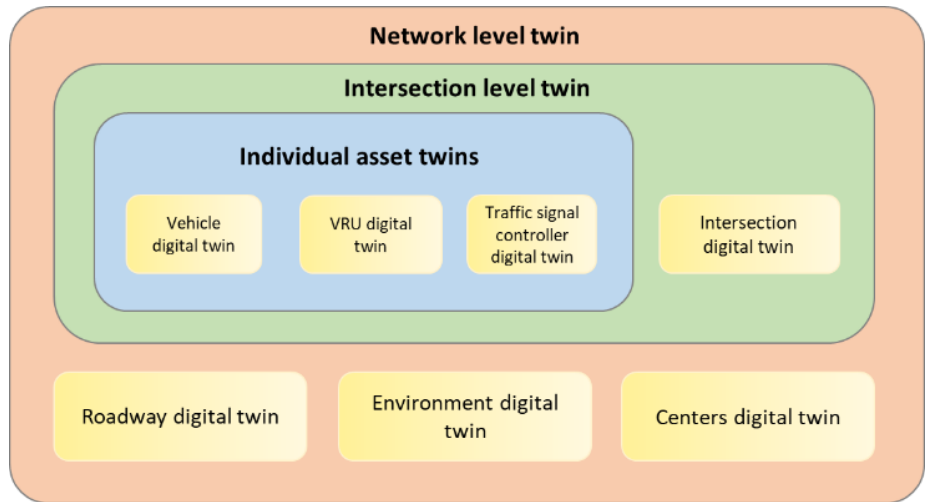

**Fig. 1.** Proposed hierarchical concept for TDT systems

The most complex and broad level within the hierarchy is the network level DT. This represents a TDT system across multiple intersections forming a network. Thus, a network level twin can be just several connected intersections or as large as a city-wide system representing the DT of each intersection and roadway section within it. Hence, it will augment the intersection level twin through the incorporation of roadway, environment and traffic center twins. Roadway twins will replicate the state of the roads and report parameters such as traffic flow, density, pavement health and any traffic incidents that may occur. For a network level twin, it is necessary to consider weather and other factors that may affect the flow of traffic. This prompts the addition of environmental state at this DT level. Under traffic centers, many different services may be grouped together including emergency and first responder services. The inclusion of centers within an DT representation of the network will enable the DT systems to coordinate signal across multiple intersections and optimize the route for such services to aid in incident or emergency response, thereby restoring mobility in the shortest duration.

In building any TDT system, it would be important to assess the requirements of the system in terms of the complexity, breadth, and depth. This will help to identify at the implementation level of the TDT system within the presented hierarchical concept. Once the level of implementation is determined the actual TDT system will need to be designed. In this regard, we propose a general reference architecture for designing TDT system keeping under consideration the requirements of the various use cases that such a TDT system may be employed towards. The fundamental parts of our proposed architecture consist of three distinct layers- the physical space, the digital space and the communication gateway layer. The synergistic coordination of the three layers enables the realization of the benefits of a TDT system. Figure 2 illustrates our proposed reference architecture for the TDT. The following sections describe each of the architecture components in detail.

*A. Physical Space*

The physical layer represents all the elements in the physical world and consists of the vehicles, Other Road Users (ORUs), traffic management infrastructure, traffic control infrastructure, roadside infrastructure, roadway equipment as well as information from the non-transportation services like weather and emergency services. The transportation related elements can be broadly group together into fixed and mobile nodes based on whether they can change their location over time. In the physical layer, the main functionalities that need to be provided are the sensing and actuation capabilities. In order to satisfy the requirements of a true digital twin, two criteria need to be ensured from the data standpoint. Firstly, the data must represent the state of the physical object. Secondly, the state information must be a live representation, with the minimum possible communication delay between the physical state and the sensed state [14]. Thus, within each component of the physical world, there needs to be a system of sensors that can satisfy the abovementioned criteria and generate data for a DT system to consume.

*Mobile nodes*

The mobile nodes of the physical transportation system comprise of vehicles and ORUs. Within vehicle systems, internal sensors measure using various internal states as well as the motion states of the vehicle. The internal states of the vehicle contain crucial information related to vehicle speed, acceleration, braking and steering angles. Other sensors can measure the proximity and speeds of nearby vehicles, ORUs and other surrounding objects in the roadway environment. Such sensors include cameras, RADARs, LIDARs, ultrasonic sensors and GPS units [16]. Whilst the sensor data from a vehicle could be used to implement a DT of itself within an in-vehicle computer [17], such a twin would not satisfy the requirements of a TDT. It is imperative that the real-time data from the vehicles be transmitted to the backend infrastructure where the transportation DT and all associated safety, mobility and environmental applications can be implemented. Data from individual vehicles is especially important for developing and implementing TDT-based traffic safety applications. Apart from the road vehicles, ORUs such as pedestrians, wheelchair users and people using non-motorized vehicles also represent mobile nodes as they can change their location over time.

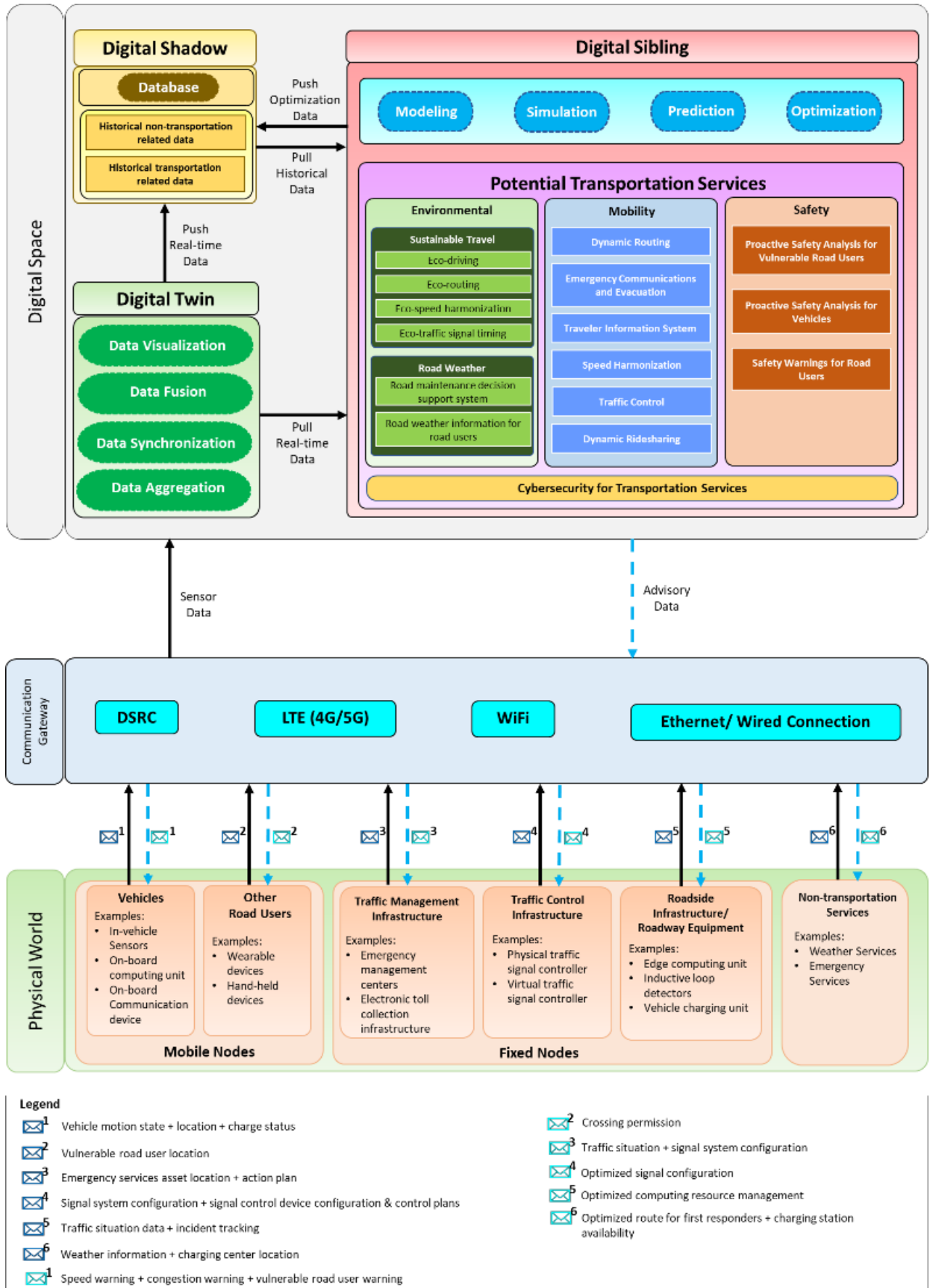

**Fig. 2.** Reference architecture for TDT.

Typically, ORUs are much slower moving as compared to vehicles, and the relevant state information for them is location information. Approximate location information of ORUs, accurate to within several meters, can be obtained from the location information transmitted by personal handheld wireless devices like smartphones.[18]. In order to obtain greater accuracy of ORU location within a specific segment of the road, roadside infrastructure such as cameras, RADARs and LIDARS may be utilized [19].

*Fixed nodes*

In contrast to the mobile nodes, the fixed nodes of the transportation system are the stationary physical assets comprising of traffic control infrastructure, roadside infrastructure, roadway equipment, and traffic management center. Traffic control infrastructure, such as traffic signal controllers, are fundamental to ensuring the mobility of transportation systems. The signal phase and timing information of the deployed signal controllers and the current updated phase information of the signal controllers will need to be monitored and replicated in the DT versions to construct the virtual representation of the physical controllers. Roadside infrastructure and roadway equipment are the traffic infrastructures placed on the side of the road and embedded within the pavement, respectively. Thus, roadside equipment will include sensors, such as cameras, LIDARs, RADARs, wireless communication radios and low-powered (i.e., edge) computing devices. These can monitor the traffic states on a roadway segment or intersection, thereby providing real-time data related to the traffic volumes and conditions on the road. Roadway equipment, such as inductive loop detectors, can also sense the roadway occupancy information. Other roadway equipment includes future wireless charging systems embedded within the pavement for wirelessly charging electric vehicles while in motion. DT can be used to optimize such a charging system and hence the data from these systems can be used to construct a DT version for monitoring and optimization.

*Non-transportation services*

The non-transportation services include all such services that are not directly part of the transportation system but can indirectly influence the mobility and safety of the system. This includes weather services and emergency services. In contrast to the other components, these services will only send the relevant information to the DT and will receive no feedback or output from the backend. Adverse weather conditions can impact mobility by causing damage to transportation infrastructure or making it hazardous to operate vehicles. Hence, it must be accounted for and replicated in the DT version of the transportation system to enable accurate analytics to be performed. In addition, emergency services like police, ambulance and first responders have very specific mobility demands due to the nature of their operation. Information from the related centers will also need to be transferred to the digital twin.

*Actuation*

Apart from the sensing function, actuation functionality is also required to be provided in the physical space. Actuation in this context can take the form of receiving and acting upon advisory messages from the digital space, such as suggested routes to avoid congestion, or it can be in the form of actual change in state of an object, for example, changing the traffic light phase. The actuation capabilities enable the TDT to implement the outputs of its safety, mobility and environmental applications.

*B. Communication gateway*

As the interface between the physical space and the digital space, the communication gateways facilitate the transmission of data between them. Data from the fixed and mobile nodes as well as non-transportation services are pushed to the digital space and actuation signals from the digital space are routed to the relevant physical asset. There are a multitude of wireless communication options that can be adopted for use within this layer. However, the DT system requires a number of requirements to be fulfilled by the communication technology that is used. Firstly, it must have a high-bandwidth and high data rate capability. Secondly, there must be minimum latency or delay in the transmission of data and information. Finally, as the system will consists of a large number of data sources, the communication technology must be able to be easily scalable as required.

Communications in DT systems can be classified into three categories: Physical to Virtual (P2V), Physical to Physical (P2P) and Virtual to Virtual (V2V) [20]. P2V communications refers to when a physical object uses wireless communication technology to communicate with a virtual twin, sharing the physical object's data in real time and receiving input from the virtual twin. Here, wide-area network wireless communication technology, such as Long Range radio (LoRa) and 5G/6G cellular communications, is mostly used. In this case, the physical object is a wireless terminal that is linked to the wireless access network via a wireless communication base station and then to the virtual twin on the Internet.

P2P communications enable information sharing between physical objects. Furthermore, a variety of communication protocols, such as the wireless personal area network (WPAN), ZigBee, and low-power wide-area networks (LPWANs) technologies such as LoRa and narrowband IoT (NB-IoT) enable network connectivity [20].

V2V communications occur in the virtual space. It imitates the communication behavior seen in the actual world. In a transportation DT context, V2V communications represent the data transmission between the DT model entities of the vehicles. Unlike real vehicle communications, which use wireless spectrum and radio power, this virtual mode relies heavily on the processing capacity of DT servers to replicate data transfer behavior [20]. Table 1 presents the properties of the available wireless communication options [21], [22].

Apart from the wireless network technology, the choice of messaging protocols is also important in digital twin. Due to the low latency requirement, lightweight messaging protocols are preferred instead of traditional http protocol for the transmission of data to the cloud. Here, two candidates are the most preferred: the Message Queue Telemetry Transport (MQTT) protocol and the Constrained Application Protocol (CoAP) [20].

TABLE I
PROPERTIES OF VARIOUS WIRELESS COMMUNICATION PROTOCOLS

| Wireless communication protocol | Frequency spectrum | Range | Data rate | Potential use case |
|---|---|---|---|---|
| LPWAN | 2.4 GHz | 25-50 m | 50 kbps | Individual level twin |
| ZigBee | Unlicensed subGHz | 10-100 m | 250 kbps | Intersection level twin |
| NB-IoT | Cellular band | 1-15 km | 250 kbps | Network level twin |
| 5G | Cellular band | Up to 100 km | 599 Mbps | Network level twin |
| LoRa | Unlicensed subGHz | 2-15 km | 50 kbps | Network level twin |

*D. Digital space*

The digital space collectively describes all the digital components of the TDT system. Real-time data from the physical space serves as the input to the digital processes and the output takes the form of control signals or advisory messages. The three main constituents for this part of the architecture are: (i) the digital twin; (ii) the digital shadow; and (iii) the digital sibling.

*Digital Twin*

The Digital Twin represents the core functionality of the TDT system architecture. Essentially this is the exact digital replica of the physical transportation assets and the related non-transportation services. State information for these physical elements transmitted to the backend or cloud systems that host all the digital elements of the TDT. A number of operations are then performed on the data in order to create the DT representations. Data from disparate sources may need to be aggregated via data aggregation operations in order to have a single source from which required data may be pulled as required. Often this aggregation operation pools data into repositories called data lakes and data warehouses. In particular, data lakes allow the storage of raw structured, semi-structured or unstructured data and have the advantage of being easily scalable to meet the demands of real-time operations [23]. Hence, for real time digital replicas, data lakes provide an ideal solution for aggregating the data. Another important operation on the data that needs to be performed is the synchronization of sensor data across multiple sensors. Within the physical space of the TDT sensors will also be spatially distributed throughout the entire transportation network. This can induce variable data transmission delays due to the differences in transmission path length. Also, it is unlikely that all the sensors and data sources will have the same sampling rate or data output rate. Thus, in order to make this data useful for real time applications, the aggregated data needs to be synchronized in a common timeframe prior to using it for state representation of an object. This will ensure that the virtual state representations are synchronized with the physical states of the object that is being replicated [24]. Data from a single sensor modality may not be adequate to determine the state of the physical object due to the inevitability of factors such as noise, occlusion, sensor failure, etc. In such instances, data fusion techniques can help to improve the precision and accuracy of a particular state reading by combining sensor data across two or more sensor modalities.

*Digital Shadow*

In the most basic form of a digital twin, the requirement is to create a real-time virtual replica of the physical asset. However, the true benefit of a DT system is realized when it can be used for both historical data analysis as well as for predictive analytics of the future state of the system. The former benefit can be obtained by maintaining historical records of the data at regular time intervals or at points of particular interest. Thus, in order to satisfy this requirement, the architecture includes a digital shadow component which acts as historical data storage for relevant transportation and non-transportation data of the TDT system. Parallel to data lakes in the DT component, data warehouses are better suited for use in the digital shadow component [23]. Data warehouses can store large amount of data which unlike data lakes, must be stored in structured form. This makes it appropriate for use as a historical data store, as it is often required to perform complex data queries on such data. Within the architecture, the DT component can push data to the digital shadow and when required to save any relevant state information at a particular time. The digital shadow can also interact with the digital sibling component of the TDT architecture. Historical data may be pulled from the digital shadow and conversely, the digital sibling can also push any data to it in order to save it.

*Digital Sibling*

The final component of the TDT architecture is the digital sibling. The digital sibling component performs all the computation tasks that are required to implement the safety, mobility and environmental applications of the TDT. Thus, the major functionalities of this component are modeling, simulation, prediction and optimization. Modeling involves constructing relevant models of the physical object, which can enable interaction with the objects in simulation environments. Thus, geometric modeling, physics-based modeling and data-driven modeling are utilized here to enable interaction with the digital replicas of the assets. Whilst the availability of real-time data in TDT system would make it easier to develop and train data-driven models, physics-based modeling can also complement them. Aside from the modeling of individual assets, it will also be necessary to model the specific transportation services that will need to be implemented. The developed models can be used to implement specific simulation scenarios of traffic environments. The simulation can use the modeled elements to observe the interaction between these elements in the digital space without the interactions ever taking place in the physical world. For example, parallel simulations can be performed to evaluate the outcomes of routing decisions and identify the best choice thereby. Each simulation will track the outcome of a particular routing decision to see the final impact on improving the mobility of the system. For safety applications, the digital sibling component can simulate interactions between ORUs and vehicles to identify potential conflicts and accident-prone zones. This enables the TDT to proactively identify safety and mobility issues and suggest measures for correction. Thus, prediction tasks are another functionality of the digital sibling component. Finally, the digital sibling can also enable the optimization of the models and simulations. Historical data related to transportation and non-transportation systems can be pulled from the digital shadow component of the architecture. Through the comparison of the prediction results with the historical data, it would be possible to optimize the models used in prediction as well as any tunable simulation parameters. This will refine the models and increase their fidelity over time enabling greater accuracy of prediction. The digital sibling can push the outputs of its training and simulation to the digital shadow to save them for future purposes as required.

## V. Transportation Safety, Mobility and Environmental Use Cases

This section presents several use cases of TDT systems based on the reference architecture. Specifically, the applications discussed pertain to improving the safety or mobility of the transportation system.

### A. Safety

Traffic safety applications aim to improve the safety aspects of transportation systems. Such applications mainly deal with three interactions between road users: Vehicle to ORU safety, Single Vehicle Safety, and Vehicle to Vehicle Safety. A TDT system can be used to implement specific applications to improve safety across all three types of interactions.

*Vehicle-to-ORU safety application*

Vehicle-to-ORU crashes constitute a major safety issue. Hence, it is important to ensure the safety of ORU in road-crossing scenarios. Typically, such crashes will involve drivers who are unaware of ORU presence or in situations where ORU may cross a road where it is prohibited to do so. A proactive safety application to address this will involve predicting possible vehicle and ORU crashes and informing the vehicle's driver and the ORU beforehand through warning messages. Information regarding approaching vehicles and ORU positions can be obtained from sensor data of their physical counterparts. The digital sibling component can track the trajectories of the vehicles and the ORUs in the digital space and predict possible crashes beforehand. A warning message can then be passed to the physical space to the vehicles as well as the ORUs. Additionally, the crash predictions for a particular road section can be stored in the digital shadow component. This information can be used to analyze and identify potential accident-prone zones and suggest rectification measures, such as optimizing pedestrian crossing signal timings.

*Single vehicle safety application*

Apart from vehicle-to-ORU crashes, single-vehicle crashes are another safety concern. Single-vehicle crashes can occur due to driver's negligence or vehicle issues, such as brake failure, tire blowout, and steering loss. Environmental factors may also be responsible, such as adverse weather, poor lighting conditions, obstacles on the roadway, and faults in the pavement. The vehicle twin can be monitored to predict the failure of critical components. Also, the driving pattern may be analyzed to identify potential driver negligence or distraction. Environmental data may be used to identify hazardous road conditions and generate advisory speeds, which can be sent to physical vehicles.

*Vehicle-to-vehicle crash avoidance application*

Vehicle-to-vehicle crashes can occur due to unsafe vehicles or unsafe driving/vehicle operation. In such instances, proactively identifying dangerous vehicles and drivers and warning others in their vicinity would help to avert such vehicle-to-vehicle crashes. Information from the vehicle digital twins can identify risky vehicles on the roads. Furthermore, in the event of crashes in the downstream portion of traffic, the TDT system can detect it through the road infrastructure twin and warn the upstream traffic to prevent pile-up crashes at the incident location due to sudden braking.

### B. Mobility

Whilst safety applications improve the safety of the road system, mobility applications aim to improve and facilitate the overall mobility of the system. TDT systems offer a great advantage in this regard as the optimization capability will allow the system to optimize traffic flow and reduce congestion. A few specific use cases are discussed here.

*Vehicle routing and signal optimization*

Within the digital sibling component of the digital space, the ability to carry out parallel simulations without affecting the real traffic flow enables the TDT to evaluate various routing decision prior to implementing them. The substantial computing power enabled by the backend system of the DT would allow such complicated simulations to be performed in real-time. Furthermore, as demonstrated by Dasgupta et al., TDT can also improve adaptive traffic signal controllers' performance [25].

*Cooperative driving*

The authors in [10] demonstrated a cooperative driving scheme which enabled vehicles to pass through the intersection safely using a DT of an intersection. Advisory messages from the intersection DT instructs the drivers to adjust their speeds such that the vehicles don't have to stop before proceeding. Since no stoppage is necessary in the scenario, there is significant reduction in travel time and energy consumption.

*Public transit allocation*

One challenge in ensuring mobility is identifying which areas suffer from a lack of transit options. Data from the road users who hail transit services can be used to identify which regions are in need of transit options, which areas are underserved, and which are redundant routes. Thus, the routing of transit services can be dynamically changed and the number of vehicles in a particular route can vary in response to demand.

*Traffic Incident management*

Incident management and mobility restoration can also be vastly streamlined via TDT systems. Data from the fixed and mobile nodes can be used to detect incidents in real time and suitable dispatch messages can be sent to the relevant agencies. The routing of emergency vehicles can be optimized within the DT and the physical traffic signal controllers can be coordinated in such a way as to provide minimal obstruction to emergency vehicles. This will enable faster response to incidents. On the other hand, the vehicles can be advised on lane or road closures and suitable detour options to provide unhindered traffic flow. Data around the time of the incident can be pulled from the digital shadow for later investigation into root causes of the incident and facilitate safety improvements and legal proceedings.

*Disaster management*

The TDT system can also be used for disaster response applications. In response to any natural or man-made disasters, the TDT can identify and advise evacuees towards the safest and quickest evacuation routes. Additionally, by monitoring the roadway conditions, the TDT can close down certain routes if any hazard is found there or if there is a failure of traffic control infrastructure.

*C. Environmental*

Apart from safety and mobility use cases, transportation systems must also be designed to improve environmental impact. Hence, TDT systems can help to monitor the real time pollution and emissions arising from particular roadway sections. Environmental use cases such as the following can be implemented through the TDT system.

*Eco-speed harmonization*

Heterogeneous traffic may cause excessive lane-changing maneuvers by the drivers. Consequently, greater acceleration and deceleration changes occur in the traffic, which adversely affects fuel consumption and emissions. As the TDT system aggregates the speed data from the vehicles on the roadway, a speed harmonization application can be run to identify an optimum uniform speed for the vehicles based on the traffic density and other conditions. Accordingly, the identified speed can be sent to the vehicles as a speed advisory message. This will ensure harmonization of the vehicle speeds such that they operate in the most fuel-efficient manner and minimize overall emissions.

*Eco-traffic signal timing*

In response to the real-time pollution data obtained from the roadways, the TDT system can optimize the traffic signal timings to ensure that overall emissions can be reduced in a transportation network. The optimized timing plan can be deployed to the physical traffic signal controllers in real time.

VI. TRANSPORTATION DIGITAL TWIN: CURRENT STATUS OF SAFETY AND MOBILITY APPLICATIONS

In this section, an in-depth review of the existing literature in the specific area of DT systems in transportation and related elements are presented. The reviewed studies in this field are summarized in Table II. Although substantial research are performed in the broader area of DT systems, the application of DT technology to transportation systems is a relatively new avenue of research. In spite of this, an attempt was made to create an extensive search of the related literature.

*A. Safety*

A study on implementing DT systems for transportation safety was cinducted by Alam and Saddik (2017); where they presented a DT-based architecture for CPS for a prototype Advanced Driver Assistance System (ADAS) [17]. Subsequent works also focused on safety use cases of TDT systems. Kumar et al. (2018) and Chen et al. (2018) demonstrated the use of DT systems in transportation safety and mobility [12][11]. Their works focused on the driver behavior and intention prediction in order to improve road safety. Later, Buechel et al. (2019) demonstrated a prototype 5G based infrastructure enabled DT of the road system [26]. Liao et al. (2022) implemented a traffic safety application for a cooperative ramp merging scenario [9]. Krämmer et al. (2022) developed an architecture for a DT representation of a roadway section to improve road safety [27]. Safety applications were also developed in other related areas as well which include advanced driver assistance systems, autonomous vehicles, vehicle safety applications [28]–[35].

*B. Mobility*

Similarly, other works have investigated mobility applications using DT systems. Dasgupta et al. (2021) developed a DT based adaptive traffic signal control to improve traffic mobility [25]. A DT-based ITS design was suggested by Rudskoy et al. (2021) to solve applications related to transportation, safety, and the environment [36]. Wang et al. (2022) used a DT of an intersection and road vehicles to enable cooperative driving [10]. Zhang et al. (2019) used DT to predict the trajectory of electric vehicles and facilitate charge scheduling of EVs [37]. Apart from these, other works have explored mobility applications from different contexts other than transportation such as smart logistics, autonomous driving and road systems [35], [38], [39]. For future connected mobility systems, vehicular networking optimization in the digital space was also investigated through the use of DT representations of physical and software defined vehicular networks [40], [41].

VII. CHALLENGES AND FUTURE RESEARCH DIRECTION IN TRANSPORTATION DIGITAL TWIN

In this section, we discuss the challenges that pertain to each component of the architecture presented in Figure 2. Additionally, we extend the discussion to include areas of future research for these components that will help to overcome the challenges.

*A. Digital Shadow*

The digital shadow component of the proposed TDT architecure necessitates the storage of a massive amount of data, which entails sophisticated, costly storage infrastructure. Along with volume, the data storage's writing and reading speeds are essential because real-time data will be streaming to and from the database.

TABLE II

SUMMARY OF EXISTING RESEARCH RELATED TO DIGITAL TWIN IN THE TRANSPORTATION FIELD

| Application type | Study | Twinned entities | Implementation | Contributions | Technologies/Tools/Strategies |
|---|---|---|---|---|---|
| Safety | Kumar et al. (2018) [12] | Roadway and vehicles | • Physical Space: Simulation<br>• Communication: Simulation<br>• Cyber Space: Simulation | Utilized a DT approach for Driver intention prediction and congestion avoidance | • Data analytics<br>• Virtual vehicle model<br>• Intelligent Transportation System (ITS) |
| | Chen et al. (2018)[11] | Driver | • Physical Space: Simulation<br>• Communication: Simulation<br>• Cyber Space: Simulation | Driver behavior prediction to support ADAS systems | • Mathematical modeling of driver behavior and environment<br>• Models sharing between connected vehicles to better predict and avoid collisions |
| | Liao et al. (2022)[9] | Driver and vehicle | • Physical Space: Real world<br>• Communication: Real world<br>• Cyber Space: Simulation | DT based Advanced Driver Assistance System (ADAS) to enable cooperative ramp merging | Field implementation using real vehicles and drivers |
| | Sahal et al. (2021)[28] | N/A | • Physical Space: Conceptual<br>• Communication: Conceptual<br>• Cyber Space: Conceptual | Proposed a concrete ledger-based collaborative DTs framework | • Blockchain<br>• Predictive analytics |
| | Ge et al. (2019) [29] | Vehicle | • Physical Space: Real world<br>• Communication: Real world<br>• Cyber Space: Simulation | Implemented a DT of an autonomous vehicle | • LTE/V2X technology<br>• Autonomous vehicles |
| | Piromalis and Kantaros (2022) [30] | Vehicle | • Physical Space: Conceptual<br>• Communication: Conceptual<br>• Cyber Space: Conceptual | Presented a discussion of DT technology application to the automotive industry | Reviewed relevant DT concepts for automotive DT application |
| | Rudskoy et al. (2021)[36] | N/A | • Physical Space: Conceptual<br>• Communication: Conceptual<br>• Cyber Space: Conceptual | Presented a DT architecture for implementing various services within ITS | ITS |
| | Buechel et al. (2019)[26] | Vehicle | • Physical Space: Real world<br>• Communication: Real world<br>• Cyber Space: Simulation | Prototyped a 5G-connected and automated vehicle for use in an infrastructure-enabled DT system of the road traffic | • 5G<br>• Vehicle sensors<br>• Human Machine Interface |
| | Almeaibed et al. (2021)[31] | Vehicle | • Physical Space: Conceptual<br>• Communication: Conceptual<br>• Cyber Space: Conceptual | Developed a framework for vehicular DT | Data driven vehicle |
| | Krämmer et al. (2022) [42] | Roadway and vehicles | • Physical Space: Real world<br>• Communication: Real world<br>• Cyber Space: Simulation | Implemented the DT of a section of highway | • Intelligent infrastructure system<br>• Multimodal sensor data |
| | Korostelkin et al. (2019)[32] | Vehicle | • Physical Space: Simulated<br>• Communication: Simulated<br>• Cyber Space: Simulated | Used a DT of a vehicle to optimize the frame mass of a vehicle for satisfying requirements of collision safety | • Optimization<br>• Finite element method |

| Application type | Study | Twinned entities | Implementation | Contributions | Technologies/Tools/Strategies |
|---|---|---|---|---|---|
| | Dygalo et al. (2020)[33] | Vehicle | • *Physical Space:* Real world<br>• *Communication:* Simulated<br>• *Cyber Space:* Simulated | Implemented DT of a braking system of a vehicle | Virtual and physical simulation technologies |
| | Liu et al. (2020)[34] | Vehicle | • *Physical Space:* Simulated<br>• *Communication:* Simulated<br>• *Cyber Space:* Simulated | Proposed a sensor fusion method to fuse in-vehicle camera data with cloud DT information | • Human-in-the-loop simulation<br>• Sensor fusion |
| | Kaliske et al. (2021)[35] | Road surface | • *Physical Space:* Conceptual<br>• *Communication:* Conceptual<br>• *Cyber Space:* Conceptual | Presented an overview of the requirements for building DT of a road system | Reviewed current teachnologies for DT of road system |
| | Wang et al. (2021)[43] | Vehicles, road networks | • *Physical Space:* Simulated<br>• *Communication:* Simulated<br>• *Cyber Space:* Simulated | Developed a DT simulation to present a case study of a personalized adaptive cruise control system | Human-in-the-loop simulation using the Unity game engine |
| | Alam and Saddik (2017)[17] | Vehicle | • *Physical Space:* Conceptual<br>• *Communication:* Conceptual<br>• *Cyber Space:* Conceptual | Presented a DT architecture for cloud-based cyber-physical system and a prototype for an advanced driver assistance system | • Cloud based cyber physical system<br>• Analytical modeling |
| Mobility and environmental* | Barosan et al. (2020)[38] | Vehicle | • *Physical Space:* Simulated<br>• *Communication:* Simulated<br>• *Cyber Space:* Simulated | Developed and implemented a DT for autonomous truck driving at distribution centers | • Virtual world using Unity game engine<br>• CARLA simulation |
| | Dasgupta et al. (2021)[25] | Traffic signal controller | • *Physical Space:* Simulated<br>• *Communication:* Simulated<br>• *Cyber Space:* Simulated | Developed a DT based Adaptive traffic signal controller to minimize traffic delay in a congested situation | • Waiting time-based adaptive traffic signal controller<br>• SUMO simulation |
| | Wang et al. (2022)[10] | Vehicle, Intersection | • *Physical Space:* Simulated<br>• *Communication:* Simulated<br>• *Cyber Space:* Simulated | Developed a DT framework for enabling cooperative driving system at a non-signalized intersection | • Agent-based modeling<br>• Human in the loop simulation |
| | Marai et al. (2020)[39] | Road infrastructure | • *Physical Space:* Real world<br>• *Communication:* Real world<br>• *Cyber Space:* Real world | Proposed method of creating DT of road infrastructure for smart cities | • Iot sensors<br>• Object detection and recognition |
| | Zhang et al. (2022)[41] | Vehicular edge network | • *Physical Space:* Real world<br>• *Communication:* Simulated<br>• *Cyber Space:* Simulated | Proposed DT-based modeling to conduct replication of the charging and discharging process of large-scale mobile EVs in different kinds of dynamic scenarios | • Multiagent deep reinforcement learning<br>• Vehicular edge computing |

* Research focusing explicitly on environmental applications of TDT system could not be found in the authors' literature search. However, studies on mobility improvements using TDT were directly associated with reducing fuel consumption and green house gas emissions, which are the objectives of environmental applications using TDT systems.

Real-time data is stored in the digital shadow, and historical data is transmitted to the digital sibling, where time-critical mobility and safety-related applications run in real-time using the digital shadow data. Structured, semi-structured, and unstructured data are generated by mobile and fixed nodes. As a result, data stored in digital shadow has Big Data characteristics, i.e., volume, velocity, and variety. Traditional storage facilities are incapable of meeting the complexity and speed demands of Big Data. It is difficult to create a dynamic Big Data storage and management solution capable of supporting networkwide TDT.

Future research should focus on building storage facilities that can satisfy the needs of massive TDT systems. Additionally, future data storage for use in the digital shadow component need to be able to serve the millions of data write and data query operations that will be generated by all of the physical components of the TDT system. Fast compression algorithms need to be developed in order to ensure data can be quickly compressed and stored. Development of database systems specifically tailored for use in TDT systems can greatly help to reduce data writing and reading speeds.

*B. Digital Sibling*

*Modeling*

Modeling is used to simulate the complex physical systems found in the real world. Because of the high level of uncertainty in human behavior, accurately modeling the physical world is extremely challenging. Two approaches can be used to model a transportation network: i) mathematical modeling or ii) data-driven modeling. For modeling purposes, mathematical models rely on physics-based equations that include certain assumptions. Due to the nonlinearity, high coupling, and time-variant nature of the models, such models cannot represent all the components and scenarios in the physical world with less uncertainty. Because the DT platform ensures the availability of a massive amount of data, data-driven models appear to be more appropriate. However, data quality and the need for a balanced dataset can significantly reduce the efficacy of data-driven models. Both types of modeling have high computational requirements that necessitate significant investment.

Future transportation research should aim to develop models for the transportation elements for a TDT system. Extensive research is required to model the behavior of ORUs in a roadway environment. In addition to model development, extensive validation of the developed models need to be performed in order to ensure universal application.

*Simulation*

Micro-mobility simulators, such as SUMO, Prescan, Vissim, and MATLAB are used to simulate various transportation scenarios. Accurate demand modeling, accurate driving behavior mimicking, and the stochastic nature of traffic are major research challenges for successful traffic simulation. Furthermore, with a large network, the simulation becomes computationally expensive, and they are not suitable for real-time safety-related applications due to the high computational time requirements. Simulating an entire road network while meeting critical time constraints is a difficult task.

Large scale simulations as is required by TDT systems, will require enormous computation power. For a highly dynamic system like transportation system, traditional parallel computing capability will fail to satisfy the simulation requirements. In this regard, a promising future alternative that is currently in development is quantum computing. Quantum computers will have the advantage of being able to run highly complex parallel operations. As viable quantum computers become a reality in the near future, it is necessary to start developing algorithms that can run traffic simulations efficiently on a quantum computing platform.

*Prediction*

Many DT decision-making applications include prediction. Traffic controller operation is optimized for serving traffic based on the predicted demand for a specific time of day for a specific roadway section. Machine learning (ML) algorithms are mostly used for prediction, which is a data-driven approach. Consequently, if the ML algorithms encounter a data point that significantly differs from the training dataset, the algorithms may output inaccurate result. This prevents the use of such ML algorithms for safety applications.

To deal with the uncertainty in traffic flow, robust models must be developed. Explainable ML models need to developed and extensively trained and tested to achieve high percentage of accuracy in order to make them suitable for use in predictive tasks for safety applications.

*Optimization*

All environmental, mobility and safety-related digital shadow applications should be designed to make the best use of both data storage and computational resources. This is more difficult to achieve because the optimization must be done in real-time. Furthermore, local applications optimize traffic, transportation assets, and special event (incident, accident, gameday, etc.) optimization tasks. Network-wide traffic flow optimization is a computationally heavy task, and doing so in real-time is even more difficult. Furthermore, traffic must be diverted via an alternate route if an accident occurs, and the best route for first responders must be calculated. The DT performs these two tasks concurrently, making them more complex.

Once again, quantum computers can have promising application in this area. Quantum computers will enable optimization of highly complex traffic optimization problems. Existing traffic optimization algorithms will need to be modified to run on quantum computing systems.

*C. Communication*

As the connecting layer between the physical and cyber-space layers, the communication layer has to deal with all the data transmission requirements of a digital twin system. With ever increasing volumes of data being generated by today's sensors and devices, it is a challenge for communication protocols to keep up with the requirements. DT systems are highly reliant upon low latency communication mediums to ensure seamless synchronization between the digital and the cyber world. A high latency would put them out of sync and defeat the purpose of the digital twin system [44]. Using high speed wired connections and low latency wireless connections like 5G can be step towards overcoming the latency challenge. In order to reduce the amount of data that is required to be transmitted to the cloud, data compression techniques are being explored. Data compression is a challenge due to the various kinds of data that is generated by different types of sensors that will be used in transportation digital twin systems. Furthermore, the messaging protocols that are used to transmit data must be lightweight to ensure minimum latency and high data rates [44], [45].

These challenges arise from the reliance on single modes of communication options and may be addressed by the use of heterogeneous wireless networking (HetNet). HetNets employ different communication options such as 5G, LTE and WiFi with seamless switchover between the various options as required. Data trasnsmission from mobile nodes within the proposed architecture is prone to data packet loss as well as loss of network coverage. Therefore, HetNets can help to resolve these issues and maintain seamless low-latency communication between the physical assets and their digital counterparts and should be investigated for use in TDT systems. Additionally, it must be ensured that during switchovers between the communication options, no security breach occurs. Hence, secure and efficient switchover algorithms needs to developed in future research work to make HetNets viable for use in TDT[46].

*D. Privacy & Cybersecurity*

Similar to any networked system, TDT systems will have a very large attack surface for cyberattacks. Data from mobile nodes must be authenticated, i.e., the identity of the mobile node must be validated before feeding to the DT, to avoid malicious data attacks. A large number of mobile nodes and their varying speeds make authentication difficult. The DT system is responsible for ensuring the confidentiality of the vehicle identity in order to protect road users' privacy and personal security. The location of the mobile nodes can be validated against multiple sources because cyber attackers can send incorrect location information to create fake congestion in order to passively manipulate traffic controllers. The communication network is difficult to secure. Each communication channel in the network has a protocol stack and layers that must be secured. Real-time and historical data are stored in the digital shadow. If this database is breached, such a breach of road users' confidentiality could result in a major security problem. Furthermore, if the database is tampered with, the digital sibling algorithms may make incorrect decisions, resulting in network-wide failure. So, protecting road users' privacy and keeping the system safe from cyber-attacks will be a major challenge in the age of TDTs.

The use of blockchain technology will need to be investigated to ensure authentication within TDT systems. Fast, robust and computationally light encryption need to be developed for use across the computation units found in vehicles. Furthermore, algorithms to detect cyber-attacks such as Denial Of Service (DOS) attacks should be important topics for future research pertinent to TDT systems.

VIII. CONCLUSION

With the ever-growing interest in DT systems across all disciplines, the field of transportation systems will also experience widespread adoption of this technology in the near future. Hence, in this paper, a hierarchical concept of a TDT system is discussed with due considerations to the required complexity, breadth and depth of the system. Additionally, a reference architecture is presented for a TDT system. The applicability of the architecture is demonstrated by presenting several use cases related to safety, mobility and environmental applications. The existing research studies in this field are reviewed. In conjunction with the reference architecture, the hierarchical concept will enable transportation planners to design TDT systems by satisfying corresponding system requirements. Despite the current advances in research, several challenges exist that hinder the development of TDT systems. These challenges range from communication requirements to cybersecurity, data storage and modeling limitations. Thus, these challenges are identified and reviewed with a view to direct future research for solving these challenges effectively and facilitating TDT system implementation. Therefore, the review is anticipated to provide a holistic and up-to-date overview of TDT systems and will benefit practitioners and future researchers to design TDT systems suited to any safety and mobility needs.

ACKNOWLEDGMENT

This material is based on a study supported by the Alabama Transportation Institute (ATI). Any opinions, findings, and conclusions or recommendations expressed in this material are those of the author(s) and do not necessarily reflect the views of the Alabama Transportation Institute (ATI).

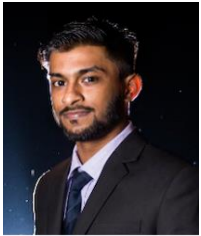
**Muhammad Sami Irfan** received the B.Sc. degree in Electrical and Electronics Engineering from the Islamic University of Technology and an MBA degree in Finance from the University of Dhaka. At present, he is a 2nd year Ph.D. student of transportation systems engineering at the University of Alabama, AL, USA. His research focuses on transportation digital twin, cybersecurity and connected and automated Vehicles.

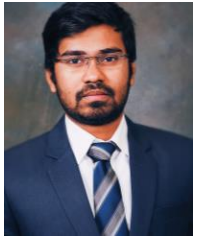
**Sagar Dasgupta** is a Ph.D student at the University of Alabama, Tuscaloosa, Alabama. He received his B.Tech. in Mechanical Engineering from Motilal Nehru National Institute of Technology Allahabad, Prayagraj, UP, India. He received his M.S. in Mechanical Engineering form Clemson University, South Carolina. His research interest includes data analytics, machine learning, connected and automated vehicles, cybersecurity, and sensor fusion.

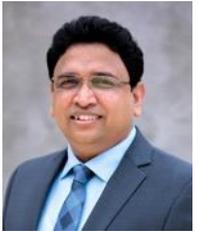
**Mizanur Rahman** is an assistant professor in the Department of Civil, Construction and Environmental Engineering at the University of Alabama, Tuscaloosa, Alabama. After his graduation in August 2018, he joined as a postdoctoral research fellow for the Center for Connected Multimodal Mobility ($C^2M^2$), a U.S. Department of Transportation Tier 1 University Transportation Center (cecas.clenson.edu/c2m2). After that, he has also served as an Assistant Director of C2M2. His research focuses on traffic flow theory, and transportation cyber-physical systems for connected and automated vehicles and smart cities.